\documentclass[print,aps,prd,superscriptaddress,nofootinbib]{revtex4-2}
\usepackage[utf8]{inputenc}
\usepackage[english]{babel}

\usepackage{amsmath,amssymb,mathtools, dcolumn, slashed, graphicx, braket, bm, xcolor, url, hyperref}
 
\DeclareMathOperator*{\sumint}{%
\mathchoice%
  {\ooalign{$\displaystyle\sum$\cr\hidewidth$\displaystyle\int$\hidewidth\cr}}
  {\ooalign{\raisebox{.14\height}{\scalebox{.7}{$\textstyle\sum$}}\cr\hidewidth$\textstyle\int$\hidewidth\cr}}
  {\ooalign{\raisebox{.2\height}{\scalebox{.6}{$\scriptstyle\sum$}}\cr$\scriptstyle\int$\cr}}
  {\ooalign{\raisebox{.2\height}{\scalebox{.6}{$\scriptstyle\sum$}}\cr$\scriptstyle\int$\cr}}
} 
 
\definecolor{persianindigo}{rgb}{0.2, 0.07, 0.48}
\definecolor{plum(traditional)}{rgb}{0.56, 0.27, 0.52}
\definecolor{purplemountainmajesty}{rgb}{0.59, 0.47, 0.71}
\definecolor{raspberryrose}{rgb}{0.7, 0.27, 0.42}
\definecolor{ruby}{rgb}{0.88, 0.07, 0.37}
\definecolor{bluebell}{rgb}{0.64, 0.64, 0.82}
\definecolor{ballblue}{rgb}{0.13, 0.67, 0.8}
\definecolor{blue(ncs)}{rgb}{0.0, 0.53, 0.74}
\definecolor{blue(pigment)}{rgb}{0.2, 0.2, 0.6}
\definecolor{forestgreen(web)}{rgb}{0.13, 0.55, 0.13}

\definecolor{pastelpink}{rgb}{0.75, 0.15, 0.49}

\hypersetup{colorlinks=true,		% false: colori base; true: personalizzazione colori
	linkcolor=blue(pigment),	          				% colore per link interni (immagini, tabelle, ...) (change box color with linkbordercolor)
	citecolor=blue(pigment),					% colore per riferimenti alla bibliografia
	filecolor=blue(pigment),							% colore per collegamenti a file
	urlcolor=blue(pigment)}							% colore per collegamenti a siti web

\def\d{{\rm d}}

\def\i-{\item[-]}

\def\1{\mathbb 1}

\allowdisplaybreaks 
\usepackage{float}
\usepackage[normalem]{ulem}
\usepackage{bbold}
\definecolor{simone_comm}{RGB}{199, 104, 10}
\definecolor{umb_comm}{RGB}{60, 176, 78}
\definecolor{franc_comm}{RGB}{150, 15, 15}
\definecolor{cris_comm}{RGB}{0, 5, 75}

\renewcommand{\d}{\mathrm{d}}

\def\nostrocostruttino#1\over#2{\mathrel{\mathop{\kern 0pt \rlap
{\hbox{$#1$}}} \hbox{\kern-.135em $#2$}}}
\def\sumint{\nostrocostruttino \sum \over {\displaystyle\int}}

\begin{document} 

\title{First insight into transverse-momentum-dependent fragmentation physics at photon-photon colliders}

\author{Simone Anedda}
\email{sanedda@dsf.unica.it
}
\affiliation{Dipartimento di Fisica, Universit\`a di Cagliari, Cittadella Universitaria, I-09042 Monserrato (CA), Italy}
\affiliation{INFN, Sezione di Cagliari, Cittadella Universitaria, I-09042 Monserrato (CA), Italy}

\author{Francesco Murgia}
\email{francesco.murgia@ca.infn.it}
\affiliation{INFN, Sezione di Cagliari, Cittadella Universitaria, I-09042 Monserrato (CA), Italy}

\author{Cristian Pisano}
\email{cristian.pisano@unica.it}
\affiliation{Dipartimento di Fisica, Universit\`a di Cagliari, Cittadella Universitaria, I-09042 Monserrato (CA), Italy}
\affiliation{INFN, Sezione di Cagliari, Cittadella Universitaria, I-09042 Monserrato (CA), Italy}

\begin{abstract}
Future planned lepton colliders, both in the circular and linear configurations, can effectively work as virtual and quasi-real photon-photon colliders and are expected to stimulate an intense physics program in the next few years. In this paper, we suggest to consider photon-photon scattering
as a useful source of information on transverse momentum dependent fragmentation functions (TMD FFs), complementing semi-inclusive deep inelastic scattering and $e^+e^-$ annihilation processes, which provide most of the present phenomenological information on TMD FFs. As a first illustrative example, we study two-hadron azimuthal asymmetries around the jet thrust-axis in the process $\ell^+\ell^-\to\gamma^* \gamma\to q\bar q\to  h_1 h_2 + X$, in which in a circular lepton collider one tagged, deeply-virtual photon scatters off
an untagged quasi-real photon, both originating from the initial lepton beams, producing inclusively an almost back-to-back light-hadron pair with large transverse momentum, in the $\gamma^*\gamma$ center of mass frame. Similar processes, in a more complicated environment due to the presence of initial hadronic states, can also be studied in ultraperipheral collisions at the LHC and the planned future hadron colliders.

\end{abstract}

\date{\today}
\maketitle

\section{Introduction}
\label{sec:intro}

Nucleon tomography and the three-dimensional structure of hadrons are nowadays among the most exciting topics in hadron physics. In this context, a relevant role is played by the so-called transverse momentum dependent partonic distribution and fragmentation functions (TMD PDFs and FFs respectively, also named TMDs collectively). Compared to the usual collinear PDFs and FFs, where the transverse component of the parton (hadron) intrinsic momenta with respect to the parent hadrons (partons) is integrated over, leading to the scale dependence of the distributions, TMDs explicitly retain their dependence on intrinsic transverse momenta. As a consequence, they account for a much richer structure involving correlations among the spin, or
polarization state, and the intrinsic motion of partons and hadrons. These correlations manifest themselves in observable spin and azimuthal asymmetries in inclusive and semi-inclusive hadronic processes.  
For an up-to-date and exhaustive review on the formalism and phenomenology of TMDs
see e.g.~Ref.~\cite{Boussarie:2023izj} and references therein.

QCD factorization theorems  within the TMD approach have been proven for three fundamental processes involving two well distinct energy scales: a perturbative one, related to a large momentum transfer in the process, and a much smaller second scale comparable to the QCD scale, $\Lambda_{\rm QCD}$, and related to the intrinsic parton motion inside hadrons~\cite{Ji:2004xq,Ji:2004wu,Collins:2011zzd,
Echevarria:2011epo}: 1) Semi-inclusive deeply inelastic scattering (SIDIS), $\ell \,p(N)\to \ell^\prime\, h + X$; 2) The Drell-Yan (DY) process for inclusive dilepton production in hadronic collisions, $A B\to \gamma^*,\,Z\to \ell^+ \ell^- + X$; 3) Two almost back-to-back hadron production in lepton-antilepton annihilations, $\ell^+\ell^-\to \gamma^*,\,Z\to h_1\, h_2 + X$, also known as semi-inclusive $\ell^+\ell^-$ annihilation (SIA).

Indeed, most of available experimental information on quark TMDs come from these three processes.
Concerning phenomenology, the full (leading and, in some cases, next-to-leading twist) structure of azimuthal dependences for particle production in (un)polarized processes 
has been derived in the TMD approach for SIDIS~\cite{Mulders:1995dh,Bacchetta:2006tn,Anselmino:2011ch} and DY~\cite{Arnold:2008kf} processes, as well as for hadron-pair SIA production~\cite{Boer:1997mf,DAlesio:2021dcx}. In particular, Refs.~\cite{Anselmino:2011ch,DAlesio:2021dcx} present, respectively for the SIDIS and SIA cases, an independent derivation within the TMD approach and the helicity formalism that will be adopted also in this paper.
Other processes, like Higgs~\cite{Boer:2011kf,Sun:2011iw}, photon pair~\cite{Qiu:2011ai}, $J/\psi$ pair~\cite{Scarpa:2019fol}, $J/\psi$+photon~\cite{denDunnen:2014kjo} and C-even quarkonium production~\cite{Boer:2012bt,Kato:2024vzt} in hadronic collisions, where the dominance of the color-singlet quarkonium formation mechanism is assumed, represent another source of valuable information on TMDs, in particular the almost unknown gluon ones. 
For these reactions TMD factorization is not fully proven yet and there are some indications for possible factorization-breaking effects for processes involving four hadrons, see e.g.~Ref.~\cite{Rogers:2010dm}.

Concerning quark TMD fragmentation functions, available information mainly comes from SIDIS and $e^+e^-$ annihilation, for which however quark flavor separation is not easy. Moreover, TMD PDFs and FFs always appear coupled in SIDIS, making phenomenology even more complicated, while in $e^+e^-$ collisions, where only fragmentation processes play a role, experimental information is presently relatively scarce. It is therefore clear that more processes and observables sensible to intrinsic parton motion effects, for which TMD factorization is guaranteed, would be very helpful in improving 
the knowledge and phenomenology of TMD fragmentation functions.

 A new class of processes for which TMD factorization is expected to hold, given the clearness of the initial electromagnetic state and the presence of final state interactions only, is inclusive hadron pair production in photon-photon collisions. These can be studied at present in ultraperipheral collisions at hadron colliders, like the Large Hadron Collider (LHC) at CERN and the Relativistic Heavy Ion Collider at BNL, although in these experiments the intricate
 hadronic environment  can hinder the study of TMD observables.
Photon-photon collisions are also actively investigated as a source of information for Higgs and heavy-boson properties and decays, and light-by-light scattering. In fact, the gamma-gamma collider operational mode is considered in all major proposals for future circular and linear lepton colliders
(for more information on photon-photon physics, see e.g.~Refs.~\cite{Budnev:1975poe,Ginzburg:1981vm,Ginzburg:1982yr,Kolanoski:1984hu, Schoeffel:2020svx}).
Therefore, in this paper we propose to consider photon-photon scattering as a future tool for gaining new complementary and clean information on quark TMD fragmentation functions. One interesting point is that in photon-photon scattering flavor separation would result more effective, as compared to SIDIS or $\ell^+\ell^-$ annihilations, since the contributions of $d$, $s$, $b$ quarks should be suppressed by a relative factor 1/16 (coming from their fractional electric charge to the fourth power) with respect to those of $u$, $c$, quarks. This would certainly help in better determining quark TMD FFs and disentangle different flavor contributions.
Another advantage is that while $\ell^+\ell^-$ colliders operate at some fixed energy scale (the center of mass (cm) energy of the two leptons), photon-photon collisions allow to vary the perturbative energy scale (related to the photon virtualities). Therefore, one can study the scale dependence of the TMD FFs in the same process and experimental setup.  

One may however wonder about the reachable luminosity in the photon-photon collision mode at lepton colliders, and the attainable cm energy. Indeed, all these aspects need to be carefully considered while developing proposals for future lepton colliders and the related detectors. At this stage, we suggest to take into account 
the possibility of undertaking a fructuous and complementary analysis of the 3D structure of hadrons and TMDs, together with the main physics cases considered.
The main motivation of this paper is in fact to stimulate the study of TMD physics at future photon-photon colliders at the time the nuclear and particle physics communities keep discussing perspectives for future large-scale hadron and lepton  colliders.
As a first illustrative example, we will consider the azimuthal distribution of a pseudoscalar hadron pair, inclusively produced in opposite hemispheres with respect to the final jet thrust axis, in photon-photon collisions, $\ell^+\ell^-\to \gamma^*\gamma\to q\bar q \to  h_1 h_2 + X$, with one deeply virtual and one quasi-real photon.
To this end, we will adopt the TMD approach at leading order and leading twist, complemented by the helicity formalism, which allows us to follow step by step, in the physical process, the role of the spin and polarization state of the particles involved.

The plan of the paper is the following: In Section~\ref{sec:kin_form} we describe the formalism adopted and the kinematics of the process; Section~\ref{sec:cross-asy} is devoted to the detailed derivation of the differential cross section of the process and the measurable azimuthal asymmetries and to a discussion of their physical content. Our final conclusions and remarks are given in section~\ref{sec:concl}.
More technical details on the kinematics, the virtual photon helicity density matrix and the hard-scattering helicity amplitudes are presented in the appendices.

\section{Kinematics and Formalism}
\label{sec:kin_form}
In this section we provide the main analytic expressions and kinematical details required to derive the differential cross section for the process
\begin{equation}
\ell^+(l_+)\,+\, \ell^-(l_-)  \to \ell^+(l_+^\prime)\,+\, h_1(P_1)\,+\,  h_2(P_2) + X\,,
\label{eq:genproc}
\end{equation}
where the four-momenta of the particles involved are shown within brackets. At leading order in the electromagnetic and strong coupling constants, $\alpha$ and $\alpha_S$ respectively, the dominant channel 
of this reaction is
\begin{equation}
\gamma_1^*(q_1)\,+\, \gamma_2(q_2) \to q(K_q) \,+\bar q(K_{\bar q}) \to h_1(P_1)\,+\,  h_2(P_2) + X\,,
\end{equation}
that is the production of a quark-antiquark pair by two-photon fusion, and their subsequent fragmention into two light unpolarized or scalar mesons (we mainly have in mind pion and kaon mesons here). In Eq.~\eqref{eq:genproc} we are considering the single-tagged configuration for a leptonic circular collider ($\ell = e,\,\mu$) where, for instance, the final antilepton 
$\ell^{\prime\,+}$ with four-momentum $l_+^\prime$ is detected and the virtuality $q_1^2 = (l_+-l^{\prime}_+)^2 \equiv -Q_1^2$ of $\gamma_1$ is known, while the final lepton $\ell^{\prime\,-}$ is undetected and $\gamma_2$ can be effectively considered as (quasi)real, $q_2^2 \simeq 0$, and described by a collinear Weizs\"acker-Williams distribution inside the parent beam lepton~\cite{vonWeizsacker:1934nji,Williams:1934ad}. For completeness, we will also consider the case in which the lepton beams can be longitudinally polarized. Furthermore, we note that the two final hadrons are produced almost back to back (in the partonic cm frame), with a large transverse momentum with respect to the $\gamma^*\gamma$ axis. 
Intrinsic transverse momentum effects in the fragmentation process, encoded in the quark TMD fragmentation functions, lead to
an observable non-collinearity of the two final hadrons around the $q$-$\bar q$ axis. This in turn manifests itself as azimuthal correlations in the two-hadron angular distribution around the jet thrust axis.

Before going into more details, some comments are in order:

a) As mentioned in the introduction, we adopt here a TMD factorization approach within the helicity formalism. The simple electromagnetic initial state should guarantee the validity of the approach as in $\ell^+\ell^-$ SIA processes. Moreover, it has been proven that TMD fragmentation functions are universal and process independent (see e.g.~Refs.~\cite{Collins:2004nx,Yuan:2007nd,Gamberg:2010uw}), so that we can consider the process in Eq.~(\ref{eq:genproc}) as a useful additional tool for TMD FF phenomenology.

b) From the theoretical point of view, the two-hadron azimuthal distribution around the jet thrust-axis is the cleanest possible observable to consider; from the experimental point of view, however, it requires a good determination of the thrust-axis that can be difficult to achieve. In fact, experimental results for SIA processes are often presented for the azimuthal distribution of one final hadron around the direction of motion of the second one.  In this paper, mainly devoted to illustrate a first application of the TMD approach in photon-photon collisions, we keep on working in the thrust-axis configuration. The formalism adopted here has been already worked out also 
for the second kinematical configuration, and could be easily implemented in our case,  see Ref.~\cite{DAlesio:2021dcx} and references therein for more details.

c) TMD evolution with the energy scale has been formulated within the Collins-Soper-Sterman (CSS) approach~\cite{Collins:1981uk,Collins:1981va,Collins:1984kg,Collins:2011zzd} and the soft collinear effective theory (SCET), see e.g.~Refs.~\cite{Echevarria:2012pw,Echevarria:2012js}.
In this paper we present results in a simplified framework valid at a fixed energy scale. Full implementation of scale evolution of TMD FFs, crucial when experimental data will become available, has been already performed, using our same framework, for the $e^+e^-\to h_1\,h_2 + X$ process in Refs.~\cite{DAlesio:2022brl,DAlesio:2023ozw} and can be directly applied to the process under study. While making the analytical expressions of the quantities considered more involved, it does not modify the general structure of the azimuthal modulations which are the main subject of this paper.

d) A possible competing contribution to our observable  comes from the gluon distribution $f_{g/\ell}(\xi)$ inside the second lepton, coupled to the hard process $\gamma^*\,g\to\,q\,\bar q $.
However, this contribution should be suppressed with respect to the photon one, since one needs first to produce a $q\bar q$ pair by a primary photon inside the lepton to generate a gluon.
This is only partially compensated by the order $\alpha\,\alpha_S(Q^2)$ of the cross section for the hard $\gamma^*\,g\to q\,\bar q$ process, as compared to the $\gamma^*\,\gamma\to q\,\bar q$ one of order $\alpha^2$.
Moreover, in principle this contribution can be  distinguished experimentally by the presence of additional hadronic production along the second lepton beam, which is suppressed in  photon-photon scattering. Anyway, the $\gamma$-gluon contribution could be easily added to the $\gamma$-$\gamma$ one by simply adapting the results presented below, implementing the required changes in the couplings and replacing the photon distribution in the second lepton, $f_{\gamma/\ell}$, by the gluon one. In this case the results are very similar to those discussed in the literature for the SIDIS case, see e.g.~Ref.~\cite{Pisano:2013cya} and references therein. In fact, we have verified that, with due changes, our results agree with those of Ref.~\cite{Pisano:2013cya} for the common parts.
 
Concerning the kinematics of the process, different cm frames enter into play: the lepton-beam $\ell^+\ell^-$ cm frame, the $\gamma_1^*\,\ell^-$ (analogous to the $\gamma^*\,p$ frame commonly adopted in SIDIS) and the $\gamma_1^*\,\gamma_2$ ones.  We will summarize in this section the derivation of the differential cross section of the process under investigation.
The azimuthal distributions of the two final hadrons around the final jet axis, which in our lowest-order (in the strong coupling constant) analysis coincides with the $q\bar q$ axis, will be discussed in the following section. Useful kinematical relations in the different cm frames, required for the calculations, are collected and discussed in Appendix~\ref{App-kin}. 

To summarize what has been said so far, we will adopt a leading order and leading twist TMD factorization approach within the helicity formalism, which clearly describes, for each step of the process, the polarization state of the particles involved and their role in the measured azimuthal distributions.
Given the simplicity of the initial (electromagnetic) state, the process under consideration is on an equal footing with direct lepton-antilepton annihilation, for which TMD factorization has been fully proven. Therefore, we are confident that the same approach can be applied to photon-photon collisions. 

Within the above described framework, the differential cross section for the process in Eq.~(\ref{eq:genproc})  can be written
as follows:
\begin{align}
 \d\sigma^{\ell^+\ell^-\to\,\ell^{\prime +}\, h_1\, h_2\, X} &=
 \frac{1}{4\,(l_+\cdot l_-)}\,\frac{\d^3\bm{l}^{\prime}_+}{2(2\pi)^3\,l^{\prime\,0}_+} \frac{\d^3\bm{K}_q}{2(2\pi)^3 K_q^{0}}\,\,\frac{\d^3\bm{K}_{\bar q}}{2(2\pi)^3 K_{\bar q}^{0}}\,(2\pi)^4\,\delta^{(4)}(q_1+q_2-K_q-K_{\bar q})\, \nonumber \\
 &\times \sum_q\,\sum_{\{\lambda_i\}}\,\tilde\rho_{\lambda_1,\lambda_1^\prime}(\gamma_1^*)\,\rho_{\lambda_2,\lambda_2^\prime}(\gamma_2)\,f_{\gamma/\ell^-,{\cal P}^{\ell^-}_{\hat z_-}}(\xi)\,\frac{\d\xi}{\xi}\,\hat{H}_{\lambda_q,\lambda_{\bar{q}};\lambda_1,\lambda_2}\,\hat{H}^*_{\lambda^\prime_q,\lambda^\prime_{\bar{q}};\lambda_1^\prime,\lambda_2^\prime}\, \label{eq:gensig} \\
 &\times \hat{D}^{h_1}_{\lambda_q,\lambda^\prime_q}(z_1,\bm{p}_{\perp 1})\,\d z_1\,\d^2\bm{p}_{\perp 1}\,
 \hat{D}^{h_2}_{\lambda_{\bar{q}},\lambda^\prime_{\bar{q}}}\,(z_2,\bm{p}_{\perp 2})\,\d z_2\,\d^2\bm{p}_{\perp 2}\,. \nonumber
\end{align}

In this equation, the first line contains the kinematical terms related to the initial flux factor, the $\ell^{\prime +}$ Lorentz invariant phase space factor (LIPS), which will be expressed in the $\ell^+\ell^-$ cm frame, the LIPS for the quark and the antiquark produced in the 
$\gamma_1^*\gamma_2$ annihilation, as well as the Dirac delta imposing momentum conservation in the hard process. The last two lines refer to the dynamical kernel of the cross section, including the hard scattering amplitudes and the parton hadronization process into the two final hadrons, through the TMD fragmentation functions. According to the factorization approach, in a reference frame where the initial photons move collinearly and in opposite directions, this kernel is given in terms of the distributions of the initial photons inside the parent leptons, the hard scattering amplitudes for the process $\gamma_1^*\,\gamma_2\to\,q\,\bar{q}$, and the fragmentation functions of the final quark and antiquark ($q$, $\bar q$) into the observed hadrons ($h_1$, $h_2$). Let us illustrate  all the ingredients entering this expression in more detail:

1) The first sum over (light) quark flavors extends to $q= u, \bar{u}, d, \bar{d}, s, \bar{s}$; it  can be generalized to include heavy ($c$, $b$) flavors;

2) In the second sum $\{\lambda_i\}$ stays for a sum over all involved helicity indices;

3) $\tilde\rho(\gamma_1^*)$ and $\rho(\gamma_2)$ are respectively the helicity density matrices of the virtual photon $\gamma_1^*$ and the quasi-real one, $\gamma_2$, whose expression will be given in the following;

4) $f_{\gamma/\ell,{\cal P}^{\ell}_{\hat z}}(\xi)$ is the Weizs\"acher-Williams distribution for the quasi-real photon inside the initial unpolarized (${\cal P}^\ell_{\hat z} = 0$) or longitudinally polarized (${\cal P}^\ell_{\hat z} = \pm 1$) $\ell^-$ lepton and $\xi$ the corresponding light-cone momentum fraction;

5) The $\hat{H}_{\lambda_q,\lambda_{\bar{q}};\lambda_1,\lambda_2}$'s are the helicity scattering amplitudes for the hard partonic process $\gamma_1^*(\lambda_1) + \gamma_2(\lambda_2) \to q(\lambda_q) +\bar{q}(\lambda_{\bar{q}})$;

6) Finally, the $\hat{D}^h_{\lambda_q,\lambda^\prime_q}(z,\bm{p_{\perp}})$'s are the transverse momentum dependent fragmentation functions encoding the soft fragmentation process of quark $q$ into the final hadron $h$ carrying a light-cone momentum fraction $z$ of the parent quark momentum and an intrinsic transverse momentum $\bm{p}_\perp$ with respect to its direction of motion.

An energy scale dependence of parton distribution and fragmentation functions is implied throughout this paper.

Let us now summarize, referring to Appendix~\ref{App-kin} for more details, some useful standard manipulations on the kinematical factors in the first line of Eq.~(\ref{eq:gensig}). We will adopt the usual invariant variables for deeply inelastic scattering,
\begin{equation}
 s = (l_+ + l_-)^2 = 2 l_+\cdot l_-\,,\quad x_B = \frac{Q^2}{2q_1\cdot l_-}\,,\quad y = \frac{q_1\cdot l_-}{l_+\cdot l_-}\,,
 \label{eq:dis-inv}
\end{equation}
with $Q^2 = -q_1^2 = x_B\,y\,s$.
Notice that we will neglect lepton, quark and hadron masses in the following. From Eq.~(\ref{eq:dis-inv}) it is easy to derive the flux factor in Eq.~(\ref{eq:gensig}),
$4\,l_+\cdot l_- = 2s$. The Lorentz-invariant phase-space for the final tagged positron can be written, in the $\ell^+$-$\ell^-$ cm frame, as follows:
\begin{equation}
\frac{\d^3\bm{l}_+^\prime}{2(2\pi)^3\,l_+^{\prime\,0}} =
 \frac{1}{16\pi^2}\,s\, y\, \d x_B\, \d y\,,
 \label{eq:lp-lips}
\end{equation}
where the angular dependence has been integrated over.
This can be seen explicitly by looking e.g.~at the expression of $l_+^\prime$ in Eq.~(\ref{eq:4momll})
and evaluating from there the Jacobian for the change of variables $\d(l_+^\prime)^1\d(l_+^\prime)^2\d(l_+^\prime)^3=|J|\,\d x_B\, \d y\,\d\phi_{\ell}$.
Concerning the 4-dimensional Dirac delta in Eq.~(\ref{eq:gensig}), we can write:
\begin{equation}
\delta^{(4)}(q_1+q_2-K_q-K_{\bar q}) = \delta(q_1^+ + q_2^+ -K_q^+ - K_{\bar q}^+)\,\delta(q_1^- + q_2^- -K_q^- - K_{\bar q}^-)\,\delta^{(2)}(-\bm{K}_{q\,T}-\bm{K}_{\bar q\, T})\,,
\label{eq:deltapmT}
\end{equation}
where we have switched temporarily to light-cone  4-vector components, $a^\mu =(a^+,a^-,\bm{a}_T)$, with $ a^{\pm} = (a^0 \pm a^3)/\sqrt{2}$.
The 2-dimensional Dirac delta on the transverse momenta fixes $\bm{K}_{q\,T} \equiv \bm{K}_T = - \bm{K}_{\bar q\,T}$.
Moreover, by using the results of Appendix~\ref{App-kin}, we find:
\begin{equation}
\delta(q_1^+ + q_2^+ -K_q^+ - K_{\bar q}^+)\,\delta(q_1^- + q_2^- -K_q^- - K_{\bar q}^-) = \frac{2}{ys}\,
\delta(1-\zeta_q-\zeta_{\bar q})\,\delta\left (\xi-x_B-\frac{\bm{K}_T^2}{\zeta_q\zeta_{\bar q}ys} \right )\,,
\label{eq:deltapm}
\end{equation}
where the additional invariants $\zeta_{q,\bar q}$ are defined as
\begin{equation}
\zeta_q = \frac{K_q \cdot l_-}{q_1 \cdot l_-}\,, \qquad \zeta_{\bar q} = \frac{K_{\bar q}\cdot l_-}{q_1\cdot l_-}\,, 
\label{eq:zetaqq-def-2}
\end{equation}
see Appendix~\ref{App-kin} for more details. The LIPS for the final quark-antiquark pair can be further manipulated:
\begin{equation}
\frac{\d^3\bm{K}_q}{2(2\pi)^3 K_q^{0}}\,\,\frac{\d^3\bm{K}_{\bar q}}{2(2\pi)^3 K_{\bar q}^{0}}\,\delta^{(2)}(-\bm{K}_{q\,T}-\bm{K}_{\bar q\, T})\, =\, \frac{1}{4(2\pi)^6}\,\frac{\d K_q^3}{K_q^0}\,\d^2\bm{K}_{T}\,
\frac{\d K_{\bar q}^3}{K_{\bar q}^0}\,.
\label{eq:lips-qq}
\end{equation}
By using again the results of Appendix~\ref{App-kin} we find that
\begin{equation}
\frac{\d K_q^3}{K_q^0} = \d\eta_q = \frac{\d \zeta_q}{\zeta_q}\,,
\label{eq:delta03}
\end{equation}
where $\eta_q$ is the quark pseudorapidity,
and similarly for $K_{\bar q}$.
Furthermore, $\d^2\bm{K}_{T}\,\,=\,K_T\,\d K_T\,\d \phi_q\,=\,(1/2)\,\d \bm{K}_T^2\,\d \phi_q\,$.

Inserting all these results into Eq.~(\ref{eq:gensig}) and collecting the constant factors we finally get:
\begin{align}
&\,\frac{\d\sigma^{\ell^+\ell^-\to\,\ell^{\prime +}\, h_1\, h_2\, X}}{\d x_B\,\d y\,\d\zeta\,\d\bm{K}_T^2\,\d \phi_q\,\d \xi\,\d z_1\,\d^2\bm{p}_{\perp\,1}\,\d z_2\,\d^2\bm{p}_{\perp 2}}  = \nonumber \\
&\, \frac{1}{2^9\,\pi^4}\,\frac{1}{\zeta(1-\zeta)\,\xi\,s} \nonumber \\
&\times \sum_q\,\sum_{\{\lambda_i\}}\,\tilde\rho_{\lambda_1,\lambda_1^\prime}(\gamma_1^*)\,\rho_{\lambda_2,\lambda_2^\prime}(\gamma_2)\,f_{\gamma/\ell^-,{\cal P}^{\ell^-}_{\hat{z}_-}}(\xi)\,\hat{H}_{\lambda_q,\lambda_{\bar{q}};\lambda_1,\lambda_2}\,\hat{H}^*_{\lambda^\prime_q,\lambda^\prime_{\bar{q}};\lambda_1^\prime,\lambda_2^\prime}\, \nonumber \\
 &\times \hat{D}^{h_1}_{\lambda_q,\lambda^\prime_q}(z_1,\bm{p}_{\perp 1})\,
 \hat{D}^{h_2}_{\lambda_{\bar{q}},\lambda^\prime_{\bar{q}}}\,(z_2,\bm{p}_{\perp 2})\,\delta\left (\xi-x_B-\frac{\bm{K}_T^2}{\zeta(1-\zeta) ys} \right )\,,
\label{eq:dsig-kern}
\end{align}
where $\zeta_{\bar q} =\zeta$, $\zeta_{q} = 1-\zeta$, and the remaining Dirac delta can be used either to fix $\xi$ in terms of $\bm{K}_T^2$ or viceversa. The various ingredients of the dynamical kernel in the above equation are described below.

The expression of the helicity density matrix for the tagged virtual photon $\gamma_1^*$ in the deeply inelastic scattering regime, $\rho_{\lambda_1,\lambda^\prime_1}(\gamma_1^*)$, properly normalized to unity, has been derived and discussed in detail, e.g., in Refs.~\cite{Schilling:1973ag,Budnev:1975poe,Anselmino:1998jv}.
For completeness, we summarize its derivation in Appendix~\ref{App-rho1}. Its expression in terms of the DIS invariants, in the photon helicity frame, where the photon moves along the $+\hat z$ axis and the leptonic $\ell^+$-$\ell^{\prime +}$ plane spans an azimuthal angle $\phi_\ell$ with respect to the $\hat x$-$\hat z$ plane, reads:
\begin{align}
& \rho(\gamma_1^*) =
 \frac{1}{2(2-y)^2} \nonumber \\
&\times \begin{pmatrix}
1+(1-y)^2 + {\cal P}^{\ell^+}_{\hat z_+}\, y(2-y) &
- e^{-i\phi_\ell}\,\sqrt{2(1-y)}\,\left[(2-y) + {\cal P}^{\ell^+}_{\hat z_+}\, y \right] &
- e^{-i2\phi_\ell}\, 2(1-y) \\
- e^{i\phi_\ell}\,\sqrt{2(1-y)}\,\left[(2-y) + {\cal P}^{\ell^+}_{\hat z_+}\, y \right] &
4(1-y) &
e^{-i\phi_\ell}\,\sqrt{2(1-y)}\,\left[(2-y) - {\cal P}^{\ell^+}_{\hat z_+}\, y \right] \\
- e^{i2\phi_\ell}\, 2(1-y) &
 e^{i\phi_\ell}\,\sqrt{2(1-y)}\,\left[(2-y) - {\cal P}^{\ell^+}_{\hat z_+}\, y \right] &
1+(1-y)^2 - {\cal P}^{\ell^+}_{\hat z_+}\, y(2-y) 
\end{pmatrix}\,,
\label{eq:rho1}
\end{align}
where ${\cal P}^{\ell^+}_{\hat z_+} = 0, \pm 1$ for unpolarized or longitudinally polarized initial leptons $\ell^+$ respectively and, without loss of generality, in the following we will take $\phi_\ell = 0$.
Notice that the normalized helicity density matrix ($\rho = \tilde\rho/{\rm Tr}[\tilde\rho]$, so that ${\rm Tr}[\rho] =1$) in Eq.~(\ref{eq:rho1}) has to be used in normalized observables related to $\d\sigma/\sigma$, while in the differential cross section $d\sigma$, Eq.~(\ref{eq:dsig-kern}), one has to reinstate the appropriate normalization factor, using $\tilde\rho = \rho\,{\rm Tr[\tilde\rho]}$, where
\begin{equation}
{\rm Tr}[\tilde\rho] \,=\,\frac{2\,e^2\,(2-y)^2}{Q^2\,y^2} \,\equiv\, \frac{2\,e^2\,(2-y)^2}{x_B\,y^3\,s}\,.
\label{eq:rho-norm}
\end{equation}

Concerning the untagged quasi-real photon $\gamma_2$, since it can only have $\lambda_2 =\pm 1$ helicities, its helicity density matrix can be effectively written as a $2\times 2$ matrix:
\begin{equation}
\rho(\gamma_2) = \frac{1}{2}\,
\begin{pmatrix}
1 + {\cal P}^{\gamma_2}_{\hat z_2} & 0 \\
0 & 1 - {\cal P}^{\gamma_2}_{\hat z_2}
\end{pmatrix}\,,
\label{eq:rho2}
\end{equation}
where
${\cal P}^{\gamma_2}_{\hat z_2}$
is the longitudinal component of the $\gamma_2$ polarization (pseudo)vector along its direction of motion,  ${\cal P}^{\gamma_2}_{\hat z_2} = \pm 1$. Notice that in our partonic reference frame $\gamma_1^*$ and $\gamma_2$ move back to back along the $\hat z$ axis, so that $\hat z_2 = \hat z_- = -\hat z_1 \equiv -\hat z$.
As a consequence of Eq.~(\ref{eq:rho2}), only two distinct combinations play a role in Eq.~(\ref{eq:dsig-kern}):
\begin{align}
\left[\,\rho_{++}(\gamma_2) + \rho_{--}(\gamma_2)\,\right]\, f_{\gamma/\ell^-,{\cal P}^{\ell^-}_{\hat{z}_-}}(\xi) &\, = \,
f_{\gamma,+/\ell^-,{\cal P}^{\ell^-}_{\hat{z}_-}}(\xi) +
f_{\gamma,-/\ell^-,{\cal P}^{\ell^-}_{\hat{z}_-}}(\xi)
\,=\,
f_{\gamma/\ell}(\xi) \,,\\
\left[\,\rho_{++}(\gamma_2) - \rho_{--}(\gamma_2)\,\right]\, f_{\gamma/\ell^-,{\cal P}^{\ell^-}_{\hat{z}_-}}(\xi) & \,= \,
{\cal P}^{\gamma_2}_{\hat{z}_2}\,f_{\gamma/\ell^-,{\cal P}^{\ell^-}_{\hat{z}_-}}(\xi) \, = \,f_{\gamma,+/\ell^-,{\cal P}^{\ell^-}_{\hat{z}_-}}(\xi)\,-\, 
f_{\gamma,-/\ell^-,{\cal P}^{\ell^-}_{\hat{z}_-}}(\xi)\,=\,
{\cal P}^{\ell^-}_{\hat z_-}\,\Delta_L f_{\gamma/\ell}(\xi)\,,
\label{eq:f2WW}
\end{align}
where $f_{\gamma/\ell}(\xi)\,=\,f_{\gamma,\pm/\ell^-,\pm}\,+\,\,f_{\gamma,\mp/\ell^-,\pm}$ and $\Delta_L f_{\gamma/\ell}(\xi)\,=\,f_{\gamma,\pm/\ell^-,\pm}\,-\,\,f_{\gamma,\mp/\ell^-,\pm}$
are respectively the unpolarized and longitudinally polarized Weizs\"aker-Williams parton distributions for $\gamma_2$ inside lepton $\ell^-$.

The next ingredient in the dynamical kernel of Eq.~(\ref{eq:dsig-kern}) are the helicity amplitudes $\hat{H}_{\lambda_q,\lambda_{\bar q}; \lambda_1,\lambda_2}$ for the hard-scattering process 
$\gamma_1^*(\lambda_1) + \gamma_2(\lambda_2) \to q(\lambda_q) + \bar{q}(\lambda_{\bar q})$. Their explicit leading order expression in the $\gamma_1^*$-$\gamma_2$ cm~frame is given in Appendix~\ref{App-H}.
Notice that since we are considering the production of light quarks, neglecting their masses, due to helicity conservation in the photon-quark vertices, the only non vanishing amplitudes have opposite values of the quark and antiquark helicities, which helps in simplifying the expression of the kernel in Eq.~(\ref{eq:dsig-kern}). Moreover, using  parity conservation, one can see that there are only six independent amplitudes.

The last step of the scattering process consists in
the independent fragmentation of the quark and the antiquark (produced exactly back to back in their cm~frame in the leading-order approach considered here, along the jet thrust-axis direction) in the final observed hadrons.
This non perturbative process is embodied into the transverse momentum dependent fragmentation functions $D^{h_1}_{\lambda_q,\lambda^\prime_q}(z_1,\bm{p}_{\perp 1})$ for $q\to h_1 + X$ (and analogously for the antiquark fragmentation). As a result of the explicit account of intrinsic transverse motion effects, the two observed hadrons are not anymore exactly back to back in the partonic cm frame. This generates possible azimuthal asymmetries in their distribution around the jet axis, which are the main subject of this study.

TMD fragmentation functions into unpolarized (or spinless) and spin 1/2 hadrons within the helicity formalism have been discussed in detail in Refs.~\cite{Anselmino:2005sh,DAlesio:2021dcx}. Here we only summarize some relations useful for the evaluation of the kernel.
TMD FFs for the process $a(s_a) \to h + X$, where $a$ is a quark or an antiquark,
can be written as
\begin{equation}
 \hat{D}^{h/a}_{\lambda_a,\lambda^\prime_a}(z,\bm{p}_\perp) = \sum_{\lambda_h} \sumint_{X,\lambda_X} \hat{{\cal D}}_{\lambda_h,\lambda_X;\lambda_a}(z,\bm{p}_\perp)
  \hat{{\cal D}}^*_{\lambda_h,\lambda_X;\lambda^\prime_a}(z,\bm{p}_\perp)\,, 
\label{eq:tmdFFgen}
\end{equation}
where the $\hat{{\cal D}}_{\lambda_h,\lambda_X;\lambda_a}(z,\bm{p}_\perp)$'s are soft, nonperturbative helicity amplitudes for the 
process $a(\lambda_a)\to h(\lambda_h)+X(\lambda_X)$ and the symbol $\sumint_{X,\lambda_X}$ stands for the helicity sum and phase-space integration for the final unobserved remnants, collectively named $X$, in the fragmentation process.
Using parity symmetry of strong interactions it is easy to see that for quark fragmentation into spinless or unpolarized final hadrons there are only two independent, leading twist TMD FFs: the unpolarized one,
\begin{equation}
 \hat{D}^{h/a}_{++}(z,\bm{p}_\perp) = \hat{D}^{h/a}_{--}(z,\bm{p}_\perp) =  D^{h}_{a}(z,p_\perp)\,,
\label{eq:FFunp}
\end{equation}
where $p_\perp = |\bm{p}_\perp|$,
and the Collins fragmentation function~\cite{Collins:1992kk}, describing the fragmentation of a transversely polarized quark into an unpolarized hadron,
\begin{equation}
 \hat{D}^{h/a}_{+-}(z,\bm{p}_\perp) = D^{h/a}_{+-}(z,p_\perp) e^{i\phi^h_a}\,,
\label{eq:FFcol}
\end{equation}
where $\phi^h_a$ is the azimuthal angle of the hadron $h$ momentum
in the fragmenting parton helicity frame. It is also easy to see that
\begin{equation}
 \hat{D}^{h/a}_{-+}(z,\bm{p}_\perp) = -\left[ \hat{D}^{h/a}_{+-}(z,\bm{p}_\perp) \right]^* = - D^{h/a}_{+-}(z,p_\perp) e^{-i\phi^h_a}\,.
\label{eq:FFmp}
\end{equation}
Common notations adopted in the literature for the $p_\perp$ dependent term of the quark Collins FF are:
\begin{equation}
 \Delta^N D^{h}_{a^\uparrow}(z,p_\perp) = \frac{2 p_\perp}{z m_h}\, H_1^{\perp, a}(z,p_\perp) = - i 2 D^{h/a}_{+-}(z,p_\perp) \,,
\label{eq:trento}
\end{equation}
with $m_h$ the mass of hadron $h$, while the $\uparrow$ arrow specifies that the quark $a$ is transversely polarized with respect to the plane containing the quark itself and the hadron. Notice also that
\begin{equation}
\int\,\d^2\bm{p}_\perp\,D^{h}_{a}(z,p_\perp) = D^{h}_{a}(z)\,,
\label{eq:FFz}
\end{equation}
the usual collinear unpolarized fragmentation function.
Moreover, for future use, we also define the lowest transverse moment of the Collins function,
\begin{align}
\int\,\d^2\bm{p}_\perp\, \Delta^N D^{h}_{a^\uparrow}(z,p_\perp) & \equiv
\int\,\d^2\bm{p}_\perp\,\frac{2 p_\perp}{z m_h}\, H_1^{\perp, a}(z,p_\perp) =
2 \pi\,\int\, \d p_\perp\,p_\perp\,
\Delta^N D^{h}_{a^\uparrow}(z,p_\perp)\nonumber \\ & =
\Delta^N D^{h}_{a^\uparrow}(z) =
4\,H_{1}^{\perp (1/2) a}(z)\,.
\label{eq:FFCollz}
\end{align}

\section{ Cross section and azimuthal asymmetries}
\label{sec:cross-asy}

Inserting Eqs.~(\ref{eq:rho1})-(\ref{eq:trento}) into Eq.~(\ref{eq:dsig-kern}) and using symmetry considerations, after some lengthy but straightforward calculations one finally finds 
the explicit expression of the differential cross section (in the sequel we will use ${\cal P}_\pm$ for ${\cal P}^{\ell^\pm}_{\hat z_\pm}$ for shortness):
\begin{align}
& \frac{\d\sigma^{\ell^+\ell^-\to\,\ell^{\prime +}\, h_1\, h_2\, X}\,\,({\cal P}_+,\,{\cal P}_-)}{\d x_B\,\d y\, \d\zeta\,\d\phi_q\, \d\xi\,\d z_1\, \d^2\bm{p}_{\perp 1}\, \d z_2\, \d^2\bm{p}_{\perp 2}} \, = \,\frac{3\,\alpha^3}{4\,\pi}\,\frac{1}{x_B\,y^2\,\xi^3\,s}\,\sum_q\,e_q^4 \nonumber \\
&\times\left\{\, \left[ A_U + {\cal P}_+{\cal P}_-\,A_L + \left(A_U^{\cos\phi_q}+ {\cal P}_+{\cal P}_-\,A_L^{\cos\phi_q}\right)\cos\phi_q + A_U^{\cos 2\phi_q}\cos 2\phi_q \right]\,D^{h_1}_q(z_1,p_{\perp 1}) D^{h_2}_{\bar q}(z_2,p_{\perp 2}) \right. \nonumber\\
&\quad + \left[\,\left( B_U^{\cos\phi_{12}} +  {\cal P}_+{\cal P}_-\,B_L^{\cos\phi_{12}}\right)\cos\phi_{12} + \left( B_U^{\cos(\phi_q-\phi_{12})} + {\cal P}_+{\cal P}_-\,B_L^{\cos(\phi_q-\phi_{12})} \right)\cos(\phi_q-\phi_{12}) \right. \nonumber\\
&\quad + \left( B_U^{\cos(\phi_q+\phi_{12})} + {\cal P}_+{\cal P}_-\,B_L^{\cos(\phi_q+\phi_{12})} \right)\cos(\phi_q+\phi_{12})
+ B_U^{\cos(2\phi_q-\phi_{12})}\cos(2\phi_q-\phi_{12}) \nonumber\\ &\quad \left. \left. +\, B_U^{\cos(2\phi_q+\phi_{12})}\cos(2\phi_q+\phi_{12}) \right]\,\Delta^N D^{h_1}_{q^\uparrow}(z_1,p_{\perp 1}) \Delta^N D^{h_2}_{\bar{q}^\uparrow}(z_2,p_{\perp 2}) \, \right\}\,,
\label{eq:dsigfin}
\end{align}
where we have used the remaining Dirac delta in Eq.~(\ref{eq:dsig-kern}) to fix $\bm{K}_T^2\,=\,\zeta(1-\zeta)(\xi-x_B)y s\,=\,\zeta(1-\zeta)\hat s$.
In Eq.~(\ref{eq:dsigfin}),
$\phi_q$ is the azimuthal angle of the scattering plane of the process $\gamma_1^*\gamma_2\to q\bar q$ with respect to the leptonic plane $\ell^+$-$\ell^{\prime\, +}$ for the tagged photon. Moreover, we have introduced the angle $\phi_{12} = \phi^{h_1}_q - \phi^{h_2}_{\bar q}$, where, as can be seen from Eq.~(\ref{eq:FFcol}), $\phi^{h_1}_q$ $(\phi^{h_2}_{\bar q})$ is the azimuthal angle of hadron $h_1(h_2)$ around the direction of motion of the parent quark(antiquark).
The subscripts $U$ and $L$ in the $A$, $B$ coefficients refer to configurations where both lepton beams are 
unpolarized, that is ${\cal P}_+\,=\,{\cal P}_- = 0$, or longitudinally polarized, ${\cal P}_+ = \pm 1$ and ${\cal P}_- =\pm 1$, respectively. Using also the results of Appendix~\ref{App-H}, the $A$ coefficients read 
\begin{align}
A_U & = 2\,\left\{\,[\,1+(1-y)^2\,]\,[\,x_B^2+(\xi-x_B)^2\,]\,\frac{1-2\zeta(1-\zeta)}{\zeta(1-\zeta)}\,+\,16\,(1-y)\,x_B(\xi-x_B)\,\right\}\,f_{\gamma/\ell}(\xi)\,,\nonumber\\
A_U^{\cos\phi_q} & = -\,8\,(2-y)\,\sqrt{1-y} \,(\xi-2 x_B)\,\sqrt{x_B(\xi-x_B)}\,\frac{1-2\zeta}{\sqrt{\zeta(1-\zeta)}}\,f_{\gamma/\ell}(\xi)\,,\nonumber\\
A_U^{\cos 2\phi_q} & = 16\,(1-y)\,x_B(\xi-x_B)\,f_{\gamma/\ell}(\xi) \,,\nonumber\\
A_L & = -\,2 y (2-y)\,\xi(\xi-2 x_B)\,\frac{1-2\zeta(1-\zeta)}{\zeta(1-\zeta)}\,\Delta_L f_{\gamma/\ell}(\xi)\,,\nonumber\\
A_L^{\cos\phi_q} & =\,8\, y \sqrt{1-y} \,\xi\,\sqrt{x_B(\xi-x_B)}\,\frac{1-2\zeta}{\sqrt{\zeta(1-\zeta)}}\,\Delta_L f_{\gamma/\ell}(\xi)\,.
\label{eq:Acoeff}
\end{align}
Similarly, for the $B$ terms we obtain
\begin{align}
B_U^{\cos\phi_{12}} & = \, \left\{\,[1+(1-y)^2]\,[\,x_B^2+(\xi-x_B)^2\,]\,-\,8\,(1-y)\,x_B(\xi-x_B)\,\right\}\,f_{\gamma/\ell}(\xi)\,,\nonumber\\
B_U^{\cos(\phi_q-\phi_{12})} & =\, -\,2\,(2-y)\,\sqrt{1-y}\,(\xi-2 x_B)\,\sqrt{x_B(\xi-x_B)}\,\sqrt{\frac{\zeta}{1-\zeta}}\,\,f_{\gamma/\ell}(\xi)\,,\nonumber\\
B_U^{\cos(\phi_q+\phi_{12})} & =\,2\,(2-y)\,\sqrt{1-y}\,(\xi-2 x_B)\,\sqrt{x_B(\xi-x_B)}\,\sqrt{\frac{1-\zeta}{\zeta}}\,\,f_{\gamma/\ell}(\xi)\,,\nonumber\\
B_U^{\cos(2\phi_q-\phi_{12})} & =\, 2\,(1-y)\,x_B(\xi-x_B)\,\frac{\zeta}{1-\zeta}\,\,f_{\gamma/\ell}(\xi)\nonumber\\
B_U^{\cos(2\phi_q+\phi_{12})} & =\, 2\,(1-y)\,x_B(\xi-x_B)\,\frac{1-\zeta}{\zeta}\,\,f_{\gamma/\ell}(\xi)\,,\nonumber\\
B_L^{\cos\phi_{12}} & = \, -\, y\,(2-y)\,\xi(\xi-2 x_B)\,\,\Delta_L f_{\gamma/\ell}(\xi)\,,\nonumber\\
B_L^{\cos(\phi_q-\phi_{12})} & = \, 2\, y\,\sqrt{1-y}\,\xi\,\sqrt{\,x_B(\xi-x_B)\,}\,\sqrt{\frac{\zeta}{1-\zeta}}\,\,\Delta_L f_{\gamma/\ell}(\xi)\,,\nonumber\\
B_L^{\cos(\phi_q+\phi_{12})} & = \,-\, 2\, y\,\sqrt{1-y}\,\xi\,\sqrt{\,x_B(\xi-x_B)\,}\,\sqrt{\frac{1-\zeta}{\zeta}}\,\,\Delta_L f_{\gamma/\ell}(\xi)\,.
\label{eq:Bcoeff}
\end{align}
In the above equations we have given all the expressions in terms of the variables $x_B$, $y$, $\zeta$ and $\xi$ in which the cross section is differential. Using the results of Appendix~\ref{App-kin} one can easily find the same quantities in terms of the partonic Mandelstam variables.
To collect more statistics one can first perform the change of variables $(\phi_q^{h_1},\phi_{\bar q}^{h_2})\,\to\,(\phi_q^{h_1},\phi_q^{h_1}-\phi_{\bar q}^{h_2}\equiv\phi_{12})$, then integrate Eq.~(\ref{eq:dsigfin}) over the azimuthal angle $\phi_q^{h_1}$ and the moduli of the intrinsic transverse momenta, $p_{\perp 1}$,  and $p_{\perp 2}$.
By using Eqs.~(\ref{eq:FFz}), (\ref{eq:FFCollz}), one finally gets:
\begin{align}
&
\frac{\d\sigma^{\ell^+\ell^-\to\,\ell^{\prime +}\, h_1\, h_2\, X}\,\,({\cal P}_+,\,{\cal P}_-)}{\d x_B\,\d y\, \d\zeta\,\d\phi_q\, \d\xi\,\d z_1\, \d z_2\, \d\phi_{12}} \, = \,\frac{3\,\alpha^3}{8\,\pi^2}\,\frac{1}{x_B\,y^2\,\xi^3\,s}\,
\sum_q\,e_q^4 \nonumber \\
&\times\left\{\, \left[ A_U + {\cal P}_+{\cal P}_-\,A_L + \left(A_U^{\cos\phi_q}+ {\cal P}_+{\cal P}_-\,A_L^{\cos\phi_q}\right)\cos\phi_q + A_U^{\cos 2\phi_q}\cos 2\phi_q \right]\,D^{h_1}_q(z_1)\, D^{h_2}_{\bar q}(z_2) \right. \nonumber\\
&\quad + \left[\,\left( B_U^{\cos\phi_{12}} +  {\cal P}_+{\cal P}_-\,B_L^{\cos\phi_{12}}\right)\cos\phi_{12} + \left( B_U^{\cos(\phi_q-\phi_{12})} + {\cal P}_+{\cal P}_-\,B_L^{\cos(\phi_q-\phi_{12})} \right)\cos(\phi_q-\phi_{12}) \right. \nonumber\\
&\quad + \left( B_U^{\cos(\phi_q+\phi_{12})} + {\cal P}_+{\cal P}_-\,B_L^{\cos(\phi_q+\phi_{12})} \right)\cos(\phi_q+\phi_{12})
+ B_U^{\cos(2\phi_q-\phi_{12})}\cos(2\phi_q-\phi_{12}) \nonumber\\ &\quad \left. \left. +\, B_U^{\cos(2\phi_q+\phi_{12})}\cos(2\phi_q+\phi_{12}) \right]\,\Delta^N D^{h_1}_{q^\uparrow}(z_1)\, \Delta^N D^{h_2}_{\bar{q}^\uparrow}(z_2) \, \right\}\,.
\label{eq:dsig-pTint}
\end{align}

{}From Eqs.~(\ref{eq:dsigfin}),~(\ref{eq:dsig-pTint}), and using the shorthand notation ${\rm d}\sigma({\cal P}_+,\,{\cal P}_-)$ for the differential cross section, we see that
\begin{align}
{\rm d}\sigma(0,0)\,&=\,
{\rm d}\sigma(1,0)\,=\,{\rm d}\sigma(0,1)\,=\,{\rm d}\sigma^{\rm unp}\,,\nonumber\\
{\rm d}\sigma(1,1)\,&=\,{\rm d}\sigma(-1,-1)\,, \label{eq:dsigpp}\\
{\rm d}\sigma(1,-1)\,&=\,{\rm d}\sigma(-1,1)\,.\nonumber
\end{align}

Additionally,
\begin{align}
{\rm d}\sigma^{\rm unp}\,&=\,\frac{1}{4}\,\bigl[\,{\rm d}\sigma(1,1)\,+\,{\rm d}\sigma(1,-1)\,+\,{\rm d}\sigma(-1,1)\,+\,{\rm d}\sigma(-1,-1)\,\bigr]\nonumber\\
&=\,\frac{1}{2}\,\bigl[\,{\rm d}\sigma(1,1)\,+\,{\rm d}\sigma(1,-1)\,\bigr]\,,\label{eq:dsUL}\\
\Delta_L\sigma &=\,{\rm d}\sigma(1,1)\,-\,{\rm d}\sigma(1,-1)\,=\,{\rm d}\sigma(-1,-1)\,-\,{\rm d}\sigma(-1,+1)\,.
\end{align}

We can therefore define the longitudinal spin asymmetry
\begin{equation}
A_{LL}\,=\,\frac{{\rm d}\sigma(1,1)\,-\,{\rm d}\sigma(1,-1)}{{\rm d}\sigma(1,1)\,+\,{\rm d}\sigma(1,-1)}\,=\,\frac{\Delta_L\sigma}{2\,{\rm d}\sigma^{\rm unp}}\,.
\label{eq:ALL}
\end{equation}

To isolate the factors associated to the different azimuthal modulations appearing in Eqs.~(\ref{eq:dsigfin}),~(\ref{eq:dsig-pTint}),~(\ref{eq:ALL}), it is common 
to define
appropriate azimuthal moments of the unpolarized cross section, ${\rm d}\sigma^{\rm unp}$, and the  longitudinal spin asymmetry $A_{LL}$.
In our case their general form can be given as follows:
\begin{equation}
\langle\,{\rm d}\sigma^{\rm unp}|\,n_q;m_{12}\,\rangle =\,2
\,\frac{\int\,\d\phi_q\,\d\phi_{12}\,{\rm d}\sigma^{\rm unp}(\phi_q,\phi_{12})\,\cos[n_q\phi_q+m_{12}\phi_{12}]}{\int\,\d\phi_q\,\d\phi_{12}\,\d\sigma^{\rm unp}(\phi_q,\phi_{12})}\,,
\label{eq:azimom1}
\end{equation}
and
\begin{equation}
\langle\,A_{LL}|\,n_q;m_{12}\,\rangle  =\,2
\,\frac{\int\,\d\phi_q\,\d\phi_{12}\,A_{LL}\,\d\sigma^{\rm unp}(\phi_q,\phi_{12})\cos[n_q\phi_q+m_{12}\phi_{12}]}{\int\,\d\phi_q\,\d\phi_{12}\,\d\sigma^{\rm unp}(\phi_q,\phi_{12})}\,,
\label{eq:azimom2}
\end{equation}
where $n_q = 0,1,2$ and $m_{12} = 0,\pm 1$.
Notice that in the case of Eq.~(\ref{eq:azimom1}) the trivial case $n_q = m_{12} =0$ will not be considered anymore.
The relevant azimuthal moments are summarized in Table~\ref{tab:mom}.

\begin{table}[h!]

    \centering
    \begin{tabular}{|c c | c c|}
    \hline
     $ \quad n_q \quad$ & $\quad m_{12} \quad$ & $\quad \langle\,\d\sigma^{\rm unp}|n_q;m_{12}\,\rangle \quad$ & $\quad \langle\,A_{LL}|n_q;m_{12}\,\rangle \quad $\\
     \hline
     $0$ & $0$ & - & $\displaystyle \frac{A_L}{A_U}$ \\
     $\pm 1$ & 0 & $\displaystyle\frac{A_U^{\cos\phi_q}}{A_U}$ &  $\displaystyle \frac{A_L^{\cos\phi_q}}{A_U}$ \\
     $\pm 2$ & 0 & $\displaystyle\frac{A_U^{\cos2\phi_q}}{A_U}$ & 0 \\
     0 & $\pm 1$ & $\displaystyle \frac{B_U^{\cos\phi_{12}}}{A_U}\,\frac{\sum_q\,e_q^4\,\Delta^N D_{q^\uparrow}^{h_1}\,\Delta^N D_{\bar q^\uparrow}^{h_2}}
{\sum_q\,e_q^4\,D_q^{h_1}\,D_{\bar q}^{h_2}}$ & $\displaystyle \frac{B_L^{\cos\phi_{12}}}{A_U}\,\frac{\sum_q\,e_q^4\,\Delta^N D_{q^\uparrow}^{h_1}\,\Delta^N D_{\bar q^\uparrow}^{h_2}}
{\sum_q\,e_q^4\,D_q^{h_1}\,D_{\bar q}^{h_2}}$ \\
1 & $\pm 1$ & $\displaystyle \frac{B_U^{\cos(\phi_q\pm\phi_{12})}}{A_U}\,\frac{\sum_q\,e_q^4\,\Delta^N D_{q^\uparrow}^{h_1}\,\Delta^N D_{\bar q^\uparrow}^{h_2}}
{\sum_q\,e_q^4\,D_q^{h_1}\,D_{\bar q}^{h_2}}$ &
 $\displaystyle \frac{B_L^{\cos(\phi_q\pm\phi_{12})}}{A_U}\,\frac{\sum_q\,e_q^4\,\Delta^N D_{q^\uparrow}^{h_1}\,\Delta^N D_{\bar q^\uparrow}^{h_2}}
{\sum_q\,e_q^4\,D_q^{h_1}\,D_{\bar q}^{h_2}}$ \\
$-1$ & $\mp 1$ & $//$ & $//$ \\
2 & $\pm 1$ &  $\displaystyle \frac{B_U^{\cos(2\phi_q\pm\phi_{12})}}{A_U}\,\frac{\sum_q\,e_q^4\,\Delta^N D_{q^\uparrow}^{h_1}\,\Delta^N D_{\bar q^\uparrow}^{h_2}}
{\sum_q\,e_q^4\,D_q^{h_1}\,D_{\bar q}^{h_2}}$ & 0 \\
$-2$ & $\mp 1$ & $//$ & $//$ \\
\hline
    \end{tabular}
    \caption{Summary of the relevant azimuthal moments of the unpolarized cross section, $\d\sigma^{\rm unp}$, and of the longitudinal azimuthal asymmetry, $A_{LL}$, according to Eqs.~(\ref{eq:azimom1}),~(\ref{eq:azimom2}).}
    \label{tab:mom}
\end{table}

For clarity, in Table~\ref{tab:mom}  we have simplified all the overall prefactors appearing in Eqs.~(\ref{eq:dsigfin}),~(\ref{eq:dsig-pTint}); this is correct if we are considering the fully differential cross section, without any integration over the kinematical variables $x_B$, $y$, and $\xi$; however, if we want to integrate over some of the variables in order to gather further statistics, we must explicitly reinstate the corresponding prefactors in the numerator and denominator of the moments.
The azimuthal moments in the first three rows of Table~\ref{tab:mom} do not depend (at fixed $z_1$ and $z_2$) on the fragmentation functions, therefore they do not provide any information on the Collins FF. However, they can be useful in testing the approach as far as the initial state of the process is concerned.
The azimuthal moments in the rest of Table~\ref{tab:mom} are the ones carrying information on the Collins FF. The complete prefactor ratios multiplying the ratio of products of (Collins and unpolarized) fragmentation functions can be separately integrated over some (or even all) of the 
other variables ($x_B$, $y$, $\zeta$, $\xi$). In doing so, the kinematical constraints described in Appendix~\ref{App-kin} must be taken into account in order to stay in the regime of validity of the approach.

Similar reasoning applies also to the more differential expression of the cross section in Eq.~(\ref{eq:dsigfin}), where the dependence on the moduli of the hadron transverse momenta with respect to  the parent quarks(antiquarks) is still explicit.

Let us also recall that the two angles entering the azimuthal moments, $\phi_q$ and $\phi_{12}$, are measured around different, well distinct, (light-cone) directions, the first one being the azimuthal angle of the $q$-$\bar q$ plane (relative to the leptonic plane) around the direction of motion of the two colliding photons in their cm frame, while the second one is the difference among the azimuthal angles of the two final hadrons around the parent quark(antiquark) direction of motion (coinciding with the jet thrust-axis in our leading order approach).

{}From the phenomenological point of view, it is important to stress that the ratio of combinations of Collins and unpolarized fragmentation functions appearing in Table~\ref{tab:mom} is the same one 
involved 
 in the $e^+e^-$ SIA process, $e^+e^-\to h_1\,h_2\,+\,X$, with the important difference that the fractional charge relative weight among different flavors is 16:1:1 in our case and 4:1:1 in the SIA case, respectively for $u$, $d$, $s$ flavors. Therefore, the analysis of the process considered here would contribute to better disentangle the flavor dependence of the Collins FF, especially for $u$ quarks. In fact, this is one of the main motivations for considering this observable in future photon-photon colliders. A second important benefit, once the already available full TMD scale evolution scheme will be implemented, is to study the scale dependence of the TMD Collins FF within the same process and experimental setup by varying the virtuality of the DIS photon.

\section{ Conclusions}
\label{sec:concl}

In this paper we have investigated the azimuthal asymmetries around the jet thrust-axis of light-hadron pairs inclusively produced in the collision of two photons. To this aim, we have adopted the helicity formalism  and assumed TMD factorization, which requires the two hadrons to be almost back to back and have large transverse momenta in the photon-photon cm frame. At the leading order approximation considered here, the reaction proceeds through the channel $\ell^+\ell^-\to\gamma^* \gamma\to q\bar q\to  h_1 h_2 + X$, where one tagged, deeply-virtual photon scatters off an untagged quasi-real photon, both of them originating from the initial lepton beams. 

The main motivation of the suggested measurements is to stimulate the study of TMD physics in photon-photon scattering at a time when future lepton colliders are starting to be discussed by the physics community.  In particular,  our proposal would certainly help in better determining quark TMD fragmentation functions, allowing for a more effective flavor separation than SIDIS or $\ell^+\ell^-$ annihilation processes, because the contributions of $d$, $s$, $b$ quarks should be suppressed by a charge weight factor of 1/16 (due to their fractional electric charge) with respect to those of $u$, $c$, quarks. Another advantage is that while $\ell^+\ell^-$ colliders operate at some fixed energy scale, that is the center of mass energy of the two leptons, in photon-photon scattering the perturbative energy scale, related to the photon virtualities, can be varied. One can therefore study the scale dependence of the TMD fragmentations function within the same process and experimental setup.

The azimuthal asymmetries discussed in the paper and their moments, summarized in Table~\ref{tab:mom}, involve (ratios of ) the quark unpolarized  and Collins TMD FFs, with different combinations of prefactors depending on the kinematics and polarization state of the initial photon-photon pair, and on the dynamics of the hard process. Studying the relative weight of these prefactors in different DIS and SIDIS kinematical configurations is of interest by itself for the validation of the approach.

The same formalism could also be applied to photon-photon scattering in linear lepton colliders, where the photons are generated by Compton back-scattering of laser beams off the initial leptons. It would also be interesting considering the production of hadron pairs formed by either two spin-1/2 particles or by a spin zero (a pion or a kaon) and a spin 1/2 (e.g.~a $\Lambda$ hyperon) particles. This should allow to study the interesting phenomenon of spontaneous $\Lambda$ polarization observed in different processes, like e.g.~proton-proton collisions, $e^+e^-$ SIA processes and specifically the role of quark flavor.

It still remains to be seen in detail whether the reachable luminosities in the photon-photon collision mode at lepton colliders will be sufficient for the proposed analysis, as well as the achievable range of center-of-mass energies. All these aspects should be carefully considered while developing proposals for future lepton colliders. To conclude, we note that similar processes, in a more complicated environment due to the presence of initial hadronic states, can also be studied in ultraperipheral collisions at the LHC and the planned future hadron colliders.

\section*{Acknowledgments}
This work is supported by the European Union ``Next Generation EU'' program through the Italian PRIN 2022 grant n.~20225ZHA7W.
\appendix

\section{ Kinematics}
\label{App-kin}

For completeness, we collect here the explicit expressions of the four-momenta of the particles involved in the process of interest, in the different reference frames considered.  We also summarize several useful kinematical relations.

\subsection{\label{app-kin-ll} The lepton-beam or Laboratory cm~frame}

We first consider the Laboratory, or $\ell^+$-$\ell^-$ cm frame, where the tagged lepton $\ell^+$ moves along the $+\hat{z}$ axis and the untagged one, $\ell^-$, in the opposite direction; for massless leptons, we have:
\begin{align}
l_+ & = \frac{\sqrt{s}}{2}\,(\,1,\,0,\,0,\,1\,)\,, \nonumber\\
l_- & =\, q_2/\xi\,= \,\frac{\sqrt{s}}{2}\,(\,1,\,0,\,0,\,-1\,)\,, \nonumber\\
l_+^\prime & = \frac{\sqrt{s}}{2}\,\Bigl(\,1-(1-x_B)y,\, 2\,\sqrt{x_B y (1-y)},\,0,\,1-(1+x_B)y\,\Bigr)\,, \nonumber\\
q_1 & = \frac{\sqrt{s}}{2}\,\Bigl(\,(1-x_B)y,\, - 2\,\sqrt{x_B y (1-y)},\,0,\,(1+x_B)y\,\Bigr)\,,
\label{eq:4momll}
\end{align}
where in the second line $\xi$ is the light-cone momentum fraction of the quasi-real photon $\gamma_2$ inside the untagged lepton $\ell^-$.
Without loss of generality, we have chosen the leptonic plane as the $x$-$z$ plane of our reference frame. If necessary, one can easily reinstate the most general dependence on the azimuthal angle of the leptonic plane, $\phi_{\ell}$, with respect to the $x$-$z$ plane of a general arbitrary frame.
Let us also remind the definition of the usual kinematical invariants adopted in deep inelastic scattering:
\begin{align}
 s & = (l_+ + l_-)^2 = 2\,l_+\cdot l_-\,,\quad Q^2 = -q_1^2 = - (l_+ - l^\prime_+)^2 = 2\,l_+\cdot l^\prime_+\,, \nonumber\\
 x_B & = \frac{Q^2}{2\,l_-\cdot q_1}\,,\quad y = \frac{l_-\cdot q_1}{l_-\cdot l_+}\,,\quad Q^2 = x_B\,y\,s\,.
\label{eq:xys}
\end{align}

By defining as $K_q$ and $K_{\bar q}$ the four-momenta of the quark and antiquark produced in the two-photon collision ($q = u, \bar u, d, \bar d, s, \bar s$), we can also introduce two additional invariants:
\begin{equation}
\zeta_q = \frac{K_q \cdot l_-}{q_1 \cdot l_-}\,, \qquad \zeta_{\bar q} = \frac{K_{\bar q}\cdot l_-}{q_1\cdot l_-}\,. 
\label{eq:zetaqq-def}
\end{equation}

Notice that we neglect light quark masses, therefore $K_q^2 = K_{\bar q}^2 = 0$.

\subsection{The virtual photon - untagged lepton
$\gamma_1^{*}$-$\ell^-$ cm~frame}\label{app-kin-gl}

This frame is the analogous of the $\gamma^*$-nucleon reference frame usually adopted in semi-inclusive deep inelastic scattering processes, with the untagged lepton $\ell^-$ and the quasi-real photon $\gamma_2$ playing respectively the role of the target nucleon and the collinear struck parton. In this frame $\gamma_1^*$ moves along the $+\hat{z}$ axis, and the untagged lepton, $\ell^-$, and $\gamma_2$ in the opposite direction.
The cm energy, usually named $W$, is given by
\begin{equation}
  W^2 = (q_1 + l_-)^2 = (1-x_B)\,y\,s = \frac{1-x_B}{x_B}\,Q^2\,.
\label{eq:W2}
\end{equation}
The four-momenta of the particles involved are:
\begin{align}
l_+ & = \frac{\sqrt{s}}{2}\,\frac{1}{\sqrt{(1-x_B)y}}\,\Bigl(\,1-x_By,\, 2\,\sqrt{x_B (1-x_B) (1-y)},\,0,\,1-2x_B+ x_By\,\Bigr) \,,\nonumber\\
l_- & = \, q_2/\xi\,= \,\frac{\sqrt{s}}{2}\,\frac{1}{\sqrt{(1-x_B)y}}\,(\,y,\,0,\,0,\,-y\,) \,,\nonumber\\
l_+^\prime & = \frac{\sqrt{s}}{2}\,\frac{1}{\sqrt{(1-x_B)y}}\,\Bigl(\,1-y+x_By,\, 2\,\sqrt{x_B (1-x_B) (1-y)},\,0,\,1-y-2x_B+ x_B y\,\Bigr) \,,\nonumber\\
q_1 & = \frac{\sqrt{s}}{2}\,\frac{1}{\sqrt{(1-x_B)y}}\,\Bigl(\,(1-2x_B) y,\,0,\,0,\,y\,\Bigr)\,.
\label{eq:4momgl}
\end{align}

The four-momenta of the final quark and antiquark can be written as:
\begin{align}
K_q & = K_T\,(\,\cosh\eta_q,\, \cos\phi_q,\,\sin\phi_q,\,\sinh\eta_q\,) \,,\nonumber\\
K_{\bar q} & = K_T\,(\,\cosh\eta_{\bar q},\, -\cos\phi_q,\,-\sin\phi_q,\,\sinh\eta_{\bar q}\,)\,,
\label{eq:k_qqbar}
\end{align}
where we have introduced the pseudo-rapidities of the quark and antiquark, $\eta_q$ and $\eta_{\bar q}$, that in the massless limit coincide with their rapidities.

It is easy then to verify that in this frame the invariants $\zeta_{q,{\bar q}}$ introduced in Eq.~(\ref{eq:zetaqq-def}) are given by:
\begin{equation}
\zeta_q  = \frac{k_T}{W}\,e^{\eta_q}\,, \qquad
\zeta_{\bar q} = \frac{k_T}{W}\,e^{\eta_{\bar q}}\,.
\label{eq:zetaqq}
\end{equation}

\subsection{ The virtual photon - quasi-real photon $\gamma_1^*$-$\gamma_2$ cm~frame}
\label{app-kin-gg}

This is the cm~frame of the hard partonic scattering process
$\gamma^*_1\gamma_2\to q\bar q$, again with $\gamma^*_1$ moving along the $+\hat{z}$ axis and the untagged lepton and $\gamma_2$ moving in the opposite direction. The final quark and antiquark are produced back to back in a plane forming an angle $\phi_q$ with the $\ell^+$-$\ell^{\prime +}$ leptonic plane, that we have assumed to be the $\hat{x}$-$\hat{z}$ plane. This frame is related to the previous one in section~\ref{app-kin-gl} by a Lorentz boost along the $\hat{z}$ axis specified  by the Lorentz factors:
\begin{equation}
\beta = \frac{1-\xi}{1-2x_B+\xi}\,,\qquad \gamma = \frac{1-2x_B+\xi}{2\,\sqrt{(1-x_B)(\xi-x_B)}}\,.
\label{eq:boost}
\end{equation}

The squared cm~energy in this frame is:
\begin{equation}
\hat{s} = (q_1+q_2)^2 = (\xi-x_B)ys\,=\frac{\xi-x_B}{x_B}\,Q^2\,,
\label{eq:hats}
\end{equation}
which also implies the constraint $\xi > x_B$. The four-momenta of the particles involved are:
\begin{align}
l_+ & = \frac{\sqrt{s}}{2}\,\frac{1}{\sqrt{(\xi-x_B)y}}\,\Bigl(\,\xi-x_B y,\, 2\,\sqrt{x_B (\xi-x_B) (1-y)},\,0,\,\xi-2x_B+x_B y\,\Bigr)\,, \nonumber\\
l_- & = \, q_2/\xi\,= \,\frac{\sqrt{s}}{2}\,\sqrt{\frac{y}{\xi-x_B}}\,\,(\,1,\,0,\,0,\,-1\,) \,,\nonumber\\
l_+^\prime & = \frac{\sqrt{s}}{2}\,\frac{1}{\sqrt{(\xi-x_B)y}}\,\Bigl(\,\xi-\xi y+x_By,\, 2\,\sqrt{x_B (\xi-x_B) (1-y)},\,0,\,\xi-\xi y-2x_B+ x_B y\,\Bigr)\,, \nonumber\\
q_1 & = \frac{\sqrt{s}}{2}\,\sqrt{\frac{y}{\xi-x_B}}\,\Bigl(\,\xi-2x_B,\,0,\,0,\,\xi\,\Bigr)\,.
\label{eq:4momgg}
\end{align}

As for the final quark-antiquark pair, we can write in general:
\begin{equation}
K_{q,\bar q} = \frac{\sqrt{\hat{s}}}{2}\,\Bigl(\,1,\,\pm \sin\theta_q\cos\phi_q,\,\pm \sin\theta_q\sin\phi_q,\,\pm \cos\theta_q\,\Bigr)\,, 
\label{eq:kqmu}
\end{equation}
where $\theta_q$ and $\phi_q$ are respectively the polar and azimuthal angle of the quark direction of motion in the $\gamma_1^*$-$\gamma_2$ cm frame.
It is again more convenient to consider the modulus of the $q$, $\bar q$ transverse momentum, $K_T$, and  their pseudorapidities 
$\hat{\eta}_{q,\bar q}$.
In fact, the first one is invariant, while the second ones are simply additive under boosts along the $\hat z$ axis.
In the $\gamma_1^*$-$\gamma_2$ frame (which is also the $q$-$\bar q$ cm frame) we clearly have $\hat{\eta}_q = - \hat{\eta}_{\bar q} = \hat{\eta}$. We can therefore write the quark and antiquark four-momenta as
\begin{equation}
K_{q,\bar q} = K_T\,\bigl(\,\cosh\hat{\eta},\,\pm \cos\phi_q,\,\pm \sin\phi_q,\,\pm \sinh\hat{\eta}\,\bigr)\,,
\label{eq:kmu2}
\end{equation}
and the transverse and longitudinal components of the quark three-momentum with respect to the $\hat z$ axis are:
\begin{equation}
K_T = \frac{\sqrt{\hat{s}}}{2}\,\frac{1}{\cosh\hat{\eta}}\,,\quad K_L = \frac{\sqrt{\hat{s}}}{2}\,\tanh\hat{\eta}\,.
\label{eq:kTkL}
\end{equation}

{}From this relation we can also see another useful property:
\begin{equation}
\xi = x_B + \frac{4 K_T^2 \cosh^2\hat{\eta}}{y s}\,.
\label{eq:xikT}
\end{equation}
Using Eq.~(\ref{eq:boost}) it is easy to see that, moving back to the $\gamma^*_1$-$\ell^-$ cm frame, by an inverse Lorentz boost  with rapidity
\begin{equation}
y_b = \frac{1}{2}\,\log\,\Bigl(\,\frac{\xi-x_B}{1-x_B}\,\Bigr) = 
\log\,\Bigl(\,\frac{\sqrt{\hat s}}{W}\,\Bigr)\,,
\label{eq:etab}
\end{equation}
the pseudorapidities of the quark and antiquark will become
\begin{equation}
\eta_q = \hat{\eta} - y_b\,,\quad \eta_{\bar q} = - \hat{\eta} -y_b\,,
\label{eq:eta-gl}
\end{equation}
and
\begin{equation}
\eta_q + \eta_{\bar q} = - 2 y_b = \log\,\Bigl(\,\frac{1-x_B}{\xi-x_B}\,\Bigr) = \log\,\Bigl(\,\frac{W^2}{\hat s}\,\Bigr)\,,\qquad \eta_q - \eta_{\bar q} = 2 \hat{\eta}\,.
\label{eq:eta-gl2}
\end{equation}

Following Ref.~\cite{Pisano:2013cya}, we can also write the invariant variables $\zeta_{q,\bar q}$ introduced in Eq.~(\ref{eq:zetaqq-def}) in terms of $\eta_{q,\bar q}$:
\begin{align}
\zeta_{\bar q} & \equiv \zeta = \frac{1}{e^{\eta_q-\eta_{\bar q}}+1} = \frac{e^{-\hat{\eta}}}{2\,\cosh\hat{\eta}} = \frac{K_T}{\sqrt{\hat s}} \,e^{-\hat \eta}\,, \nonumber\\
\zeta_q & \equiv 1 - \zeta = \frac{1}{e^{\eta_{\bar q}-\eta_q}+1} = \frac{e^{\hat{\eta}}}{2\,\cosh\hat{\eta}} = \frac{K_T}{\sqrt{\hat s}} \,e^{\hat \eta}\,,
\label{eq:z12}
\end{align}
{}from which it is also clear that $\zeta_q + \zeta_{\bar q} = 1$\,, and in the last equality we have used Eq.~(\ref{eq:kTkL}).

Using these results, we can also write
\begin{equation}
K_{q,\bar q} =
\frac{1}{2}\,\sqrt{\,(\xi-x_B)\,y\,s\,}\,\,\Bigl(\,1,\,\pm\, 2\,\sqrt{\,\zeta(1-\zeta)\,}\cos\phi_q,\,\pm\, 2\,\sqrt{\,\zeta(1-\zeta)\,}\sin\phi_q,\,\pm\, (1-2\zeta)\,\Bigr)\,.
\label{eq:kmu2-z}
\end{equation}
It is also useful to give the relation between these invariant variables and the Mandelstam invariants  in the two-photon cm~frame:
\begin{equation}
\hat{s} = (\xi-x_B) y s\,, \qquad \hat{t} = - \zeta\xi y s\,, \qquad \hat{u} = - (1-\zeta) \xi y s \,.
\label{eq:stu-z}
\end{equation}

Let us also remind that there are a few additional kinematical constraints that must be fulfilled in order to guarantee the validity of the formalism adopted: 1) First of all, to stay in the deeply virtual and factorization regime we require $Q^2 \geq Q_0^2 \gg \Lambda_{\rm QCD}^2$,
with $Q_0^2$ an arbitrary fixed hard scale of at least a few GeV$^2$;
2) Secondly, in order to clearly distinguish the two light-cone directions considered, the $\gamma^*\gamma$ axis and the jet thrust axis, and keep staying in the validity regime of the factorization approach adopted, we also require $|\bm{K}_{q T}| = K_T \geq Q_0$.
{}From Eq.~(\ref{eq:z12}) it is also easy to see that
\begin{equation}
K_T^2 = \zeta(1-\zeta)\hat s\,.
\label{eq:KT-zs}
\end{equation}
Therefore, putting a lower limit on the value of $K_T$, as required for the applicability of our factorization scheme, implies a lower cut on $\hat s$ and a limited range of allowed values for $\zeta$:~Since ${\rm Max}[\zeta(1-\zeta)] = 1/4$,
$\hat s \geq 4\,Q_0^2$ and, at fixed $\hat s$,
\begin{equation}
\frac{1}{2}\,\Bigl\{\,1-\sqrt{1-\frac{4\,Q_0^2}{\hat s}}\,\Bigr\}\, \leq \,\zeta\, \leq \,\frac{1}{2}\,\Bigl\{\,1+\sqrt{1-\frac{4\,Q_0^2}{\hat s}}\,\Bigr\}\,.
\label{eq:zeta-bound}
\end{equation}
For a generic $Q^2 \geq Q_0^2$, requiring that $\hat s = (\xi-x_B)Q^2/x_B \geq 4Q_0^2$ implies
\begin{equation}
\xi \geq x_B\,\left(\,1+\frac{4Q_0^2}{Q^2}\,\right)\,.
\label{eq:xigeq}
\end{equation}

Since $\xi \leq 1$, one finds
\begin{equation}
 x_B\,\leq\,x_B^{\rm max}(Q^2) \,=\,\frac{Q^2}{Q^2+Q_0^2}\,<\,1\,.
 \label{eq:xBmax}
\end{equation}

Given that $x_B^{\rm min}(Q^2)\,=\,Q^2/s$ (for $y\,=\,1$), imposing that $x_B^{\rm min}(Q^2) \,\leq\,x_B^{\rm max}(Q^2)$, one also finds an upper limit for $Q^2$, $Q^2 \leq s- 4Q_0^2$.
Furthermore,
\begin{equation}
 y^{\rm min}(Q^2)\,=\,\frac{Q^2}{x_B^{\rm max}(Q^2)\,s}\,=\,\frac{Q^2+4Q_0^2}{s}\,,
 \label{eq:ymin}
\end{equation}
and, from Eq.~(\ref{eq:xigeq}),
\begin{equation}
 \xi^{\rm min}(Q^2) \,=\,x_B^{\rm min}(Q^2)\,\left(\,1+\frac{4Q_0^2}{Q^2}\,\right)\,=\,y^{\rm min}(Q^2)\,.
 \label{eq:ximin}
\end{equation}

Summarizing, then, we have the constraints:
\begin{align}
& Q_0^2\,\leq\,Q^2\,\leq\,s-4Q_0^2\,, \nonumber\\
& \frac{Q^2}{s}\,\leq\,x_B\,\leq\,\frac{Q^2}{Q^2+4Q_0^2}\,,
\label{eq:limfin}\\
& \frac{Q^2+4Q_0^2}{s}\,\leq\,y,\,\xi\,\leq\,1\,.\nonumber
\end{align}

\section{The virtual photon helicity density matrix in DIS processes
\label{App-rho1}}

The derivation of the virtual photon helicity density matrix was given in detail, although with a notation different from the one adopted in this paper, in Ref.~\cite{Schilling:1973ag}. It was also briefly discussed, within the same notation adopted here, in Ref.~\cite{Anselmino:1998jv}.
We therefore believe it can be useful to summarize here the main steps leading to the explicit expression of Eq.~(\ref{eq:rho1}).

Let us start from the helicity amplitudes
for a generic process in which a lepton $\ell$ interacts with a given initial system $M$ by exchanging a single virtual photon, $\ell(l,\lambda_{\ell}) + M(P_i,\Lambda_i)\,\to\,\ell^\prime(l^\prime,\lambda_{\ell^\prime}) + M(P_f,\Lambda_f)$. For shortness,
$P_{i,f},\,\Lambda_{i,f}$ represent the full set of initial and final moments and helicities of $M$. 
This system can be e.g.~a second lepton, like in our case, or a proton/nucleon target like in SIDIS.

The helicity amplitude for this process can be written as follows:
\begin{align}
H_{\lambda_{\ell^\prime},\,\Lambda_f\,;\,\lambda_{\ell},\,\Lambda_i} & = e\,\bar{u}_{\lambda_{\ell^\prime}}\,\gamma^\mu\,u_{\lambda_{\ell}}\,\bigl(\,-i\frac{g_{\mu\nu}}{q^2}\,\bigr)\,M^\nu_{\Lambda_f;\Lambda_i}(P_f,P_i) \nonumber\\
& = e\,\bar{u}_{\lambda_{\ell^\prime}}\,\gamma^\mu\,u_{\lambda_{\ell}}\,\bigl[\,-i\frac{g_{\mu\nu}}{q^2}\,+\,\frac{q_\mu q_\nu}{q^4}\,\bigr]\,M^\nu_{\Lambda_f;\Lambda_i}(P_f,P_i) \label{eq:B-Hin}\\
& = e\,\frac{1}{q^2}\bar{u}_{\lambda_{\ell^\prime}}\,\gamma^\mu\,u_{\lambda_{\ell}}\,\Bigl[\,\sum_{\lambda_\gamma}\,(-1)^{\lambda_\gamma+1}\,\epsilon_{\mu,\,\lambda_\gamma}^*(q)\,\epsilon_{\nu,\,\lambda_\gamma}(q)\,\Bigr]\,M^\nu_{\Lambda_f;\Lambda_i}(P_f,P_i)\,,\nonumber
\end{align}
where in the second line we have used Ward identity and in the last line well-known properties of the polarization vectors of a virtual photon.

At the same time, we can separately define the helicity amplitudes for the processes $\ell(l,\lambda_\ell)\to \ell^\prime(l^\prime,\lambda_{\ell^\prime}) + \gamma^*(q,\lambda_\gamma)$,
\begin{equation}
{\cal H}_{\lambda_{\ell^\prime},\lambda_\gamma;\lambda_\ell}\, = \, e\,\bar{u}_{\lambda_{\ell^\prime}}\,\gamma^\mu\,u_{\lambda_{\ell}}\,\epsilon_{\mu,\,\lambda_\gamma}^*(q)\,,
\label{eq:B-calHl}
\end{equation}
and $\gamma^*(q,\lambda_\gamma) + M(P_i,\Lambda_i)\,\to\,M(P_f,\Lambda_f)$,
\begin{equation}
{\cal H}_{\Lambda_f;\lambda_\gamma,\Lambda_i}\,=\,\epsilon_{\nu,\,\lambda_\gamma}(q)\,
M^\nu_{\Lambda_f;\Lambda_i}(P_f,P_i)\,.
\label{eq:B-calHM}
\end{equation}

Let us now consider the differential cross section for the full process     
$\ell(l,\lambda_{\ell}) + M(P_i,\Lambda_i)\,\to\,\ell^\prime(l^\prime,\lambda_{\ell^\prime}) + M(P_f,\Lambda_f)$.
Assuming in general that the initial lepton $\ell$ 
is in a polarization state described by its helicity density matrix $\rho^{\ell(s_\ell)}_{\lambda_\ell,\lambda^\prime_\ell}$, where $s_{\ell}$ specifies the lepton $\ell$ spin state, and using Eqs.~(\ref{eq:B-Hin}), (\ref{eq:B-calHl}), (\ref{eq:B-calHM}), we can write
\begin{align}
{\rm d}\sigma\, & \propto\,\sum_{\lambda_\ell,\lambda^\prime_\ell,\lambda_{\ell^\prime}}\,\sum_{\Lambda_i,\Lambda_f}\,\rho^{\ell(s_\ell)}_{\lambda_\ell,\lambda^\prime_\ell}\,H_{\lambda_{\ell^\prime},\,\Lambda_f\,;\,\lambda_{\ell},\,\Lambda_i}\,H^*_{\lambda_{\ell^\prime},\,\Lambda_f\,;\,\lambda^\prime_{\ell},\,\Lambda_i} \nonumber\\
& = \sum_{\lambda_\ell,\lambda^\prime_\ell,\lambda_{\ell^\prime}}\,\sum_{\Lambda_i,\Lambda_f}\,\rho^{\ell(s_\ell)}_{\lambda_\ell,\lambda^\prime_\ell}\,\Bigl[\,\frac{1}{q^2}\,\sum_{\lambda_\gamma}\,(-1)^{\lambda_\gamma+1}\,{\cal H}_{\lambda_{\ell^\prime},\lambda_\gamma;\lambda_\ell}\,{\cal H}_{\Lambda_f;\lambda_\gamma,\Lambda_i}\,\Bigr] \nonumber\\
& \times\,\Bigl[\,\frac{1}{q^2}\,\sum_{\lambda^\prime_\gamma}\,(-1)^{\lambda^\prime_\gamma+1}\,{\cal H}^*_{\lambda_{\ell^\prime},\lambda^\prime_\gamma;\lambda^\prime_\ell}\,{\cal H}^*_{\Lambda_f;\lambda^\prime_\gamma,\Lambda_i}\,\Bigr] \label{eq:B-dsig} \\
& =\,\sum_{\lambda_\gamma,\lambda^\prime_\gamma}\,\tilde{\rho}_{\lambda_\gamma,\lambda^\prime_\gamma}(\gamma^*)\,\sum_{\Lambda_i,\Lambda_f}\,{\cal H}_{\Lambda_f,\lambda_\gamma;\Lambda_i}\,{\cal H}^*_{\Lambda_f,\lambda^\prime_\gamma;\Lambda_i}\,\nonumber
\end{align}
where we have defined the (non normalized) helicity density matrix for the virtual photon $\gamma^*$,
\begin{equation}
\tilde{\rho}_{\lambda_\gamma,\lambda^\prime_\gamma}(\gamma^*)\,=\,\frac{1}{q^4}\,(-1)^{\lambda_\gamma+\lambda^\prime_\gamma}\,\sum_{\lambda_\ell,\lambda^\prime_\ell,\lambda_{\ell^\prime}}\,\rho^{\ell(s_\ell)}_{\lambda_\ell,\lambda^\prime_\ell}\,{\cal H}_{\lambda_{\ell^\prime},\lambda_\gamma;\lambda_\ell}\,{\cal H}^*_{\lambda_{\ell^\prime},\lambda^\prime_\gamma;\lambda^\prime_\ell}\,.
\label{eq:B-helmat}
\end{equation}

This is the expression we were looking for.

The helicity density matrix for the initial lepton can be written in general as
\begin{equation}
\rho^{\ell(s_\ell)}_{\lambda_\ell,\lambda^\prime_\ell}\,=\,\frac{1}{2}\,\left(\,I + \bm{{\cal P}}^\ell \cdot \bm{\sigma}\,\right)\,=\,\frac{1}{2}\,\begin{pmatrix}
1 + {\cal P}^{\ell}_{\hat z} &  {\cal P}^{\ell}_{\hat x} +i{\cal P}^{\ell}_{\hat y}\\
 {\cal P}^{\ell}_{\hat x} -i{\cal P}^{\ell}_{\hat y} & 1 - {\cal P}^{\ell}_{\hat z}
\end{pmatrix}\,,
\label{eq:B-rhol}
\end{equation}
where $\bm{{\cal P}}^\ell$ is the polarization (pseudo)vector of the lepton.
In this paper, we will only consider the case of unpolarized or fully longitudinally polarized lepton beams (${\cal P}^{\ell}_{\hat z}= 0$ or  $\pm 1$ respectively), taking ${\cal P}^{\ell}_{\hat x} = {\cal P}^{\ell}_{\hat y} = 0$.

Inserting Eqs.~(\ref{eq:B-calHl}), (\ref{eq:B-rhol}) into Eq.~(\ref{eq:B-helmat}) and performing some traces we finally get:
\begin{align}
\tilde{\rho}_{\lambda_\gamma,\lambda^\prime_\gamma}(\gamma^*)\,=\,\frac{e^2}{q^4}\,(-1)^{\lambda_\gamma+\lambda^\prime_\gamma}\,& \left[\,4\,l\cdot\epsilon^*_{\lambda_\gamma}(q)\,l\cdot\epsilon_{\lambda^\prime_\gamma}(q)\, + q^2\,\epsilon^*_{\lambda_\gamma}(q)\cdot \epsilon_{\lambda^\prime_\gamma}(q)\, \,\right. \nonumber\\
& \left.\, +2\,i\,{\cal P}^\ell_{\hat z}\,\epsilon^{\alpha\beta\mu\nu}\,l_\alpha\,\epsilon_{\beta,\lambda_\gamma}(q)\,l_\mu\, \epsilon_{\nu,\lambda^\prime_\gamma}(q)\,\right]\,.
\label{eq:B-rhot-fin}
\end{align}

By assuming that the virtual photon moves along the positive $\hat z$ axis of our reference frame, and the leptonic $\ell$-$\ell^\prime$ plane forms an angle $\phi_\ell$ with the $\hat x$-$\hat z$ plane, and using the results of Appendix~\ref{App-kin} in the $\gamma^*$-$\ell_2$ frame,
we get:
\begin{align}
l & = \frac{\sqrt{s}}{2}\,\frac{1}{\sqrt{(1-x_B)y}}\,\Bigl(\,1-x_By,\, 2\,\sqrt{x_B (1-x_B) (1-y)}\,\cos\phi_\ell,\,2\,\sqrt{x_B (1-x_B) (1-y)}\,\sin\phi_\ell,\,1-2x_B+ x_By\,\Bigr) \nonumber\\
q & = \frac{\sqrt{s}}{2}\,\frac{1}{\sqrt{(1-x_B)y}}\,\Bigl(\,(1-2x_B) y,\,0,\,0,\,y\,\Bigr)\,.
\label{eq:B-lqkin}
\end{align}

Furthermore, one has
\begin{align}
\epsilon_{\lambda_\gamma=\pm 1}(q)\,& =\,\frac{1}{\sqrt{2}}\,\left(\,0,\,-\lambda_\gamma,\,-i,\,0\,\right) \nonumber\\ \epsilon_{\lambda_\gamma=0}(q)\,& =\,\frac{1}{2}\,\frac{1}{\sqrt{x_B(1-x_B)}}\,\left(\,1,0,0,1-2x_B\right)\,.
\label{eq:B-epsi} 
\end{align}

Inserting these expressions into Eq.~(\ref{eq:B-rhot-fin}),
after some straightforward calculation we finally get the result of Eq.~(\ref{eq:rho-norm}).

Notice that the normalized helicity density matrix of the virtual photon is entirely given in terms of the DIS invariant $y$. The only kinematical frame-dependent quantity is the azimuthal angle $\phi_\ell$ of the leptonic plane, which is however invariant under boosts along the $\hat z$ axis.

Let us finally observe that the virtual photon helicity density matrix is directly related to the usual leptonic tensor $L^{\mu\nu}$.

\section{Helicity amplitudes for the process $\gamma_1^*(\lambda_1) \gamma_2(\lambda_2)\to q(\lambda_q) \bar q(\lambda_{\bar q})$}
\label{App-H}

We list here for reference the non-vanishing helicity amplitudes $\hat{H}_{\lambda_q,\lambda_{\bar q};\lambda_1,\lambda_2}$ for the process $\gamma_1^*(\lambda_1) \gamma_2(\lambda_2)\to q(\lambda_q) \bar q(\lambda_{\bar q})$, for real photon $\gamma_2$ and massless quarks, in terms of $Q^2$, the virtuality of photon 1, the partonic Mandelstam invariants $\hat{s}$, $\hat{t}$, $\hat{u}$ 
and the azimuthal angle $\phi_q$ between the $q$-$\bar q$ plane and the leptonic plane for the virtual photon 1 (assumed to be, without loss of generality, the $\hat x$-$\hat z$ plane).  
We also give their expression in terms of the invariants $x_B$, $\xi$ and $\zeta$, that have been used in order to evaluate the $A_{U,L}$ and $B_{U,L}$ factors in Eqs.~(\ref{eq:Acoeff}) and (\ref{eq:Bcoeff}) respectively.

\begin{align}
\hat{H}_{+-;1,1} & = - \hat{H}_{-+;-1,-1}   =  - 2\,\sqrt{3}\,e^2\,e_q^2\,\frac{Q^2}{\hat{s}+Q^2}\,\sqrt{\frac{\hat{u}}{\hat{t}}}  =  -2\,\sqrt{3}\,e^2\,e_q^2\,\frac{x_B}{\xi}\,\sqrt{\,\frac{1-\zeta}{\zeta}\,}\,, \nonumber\\
\hat{H}_{+-;1,-1} & = - \hat{H}^*_{-+;-1,1}   =  -2\,\sqrt{3}\,e^2\,e_q^2\,e^{i2\phi_q}\,\frac{\hat{s}}{\hat{s}+Q^2}\,\sqrt{\frac{\hat{u}}{\hat{t}}}  = \,-2\,\sqrt{3}\,e^2\,e_q^2\,e^{i2\phi_q}\,\frac{\xi-x_B}{\xi}\,\sqrt{\,\frac{1-\zeta}{\zeta}\,}\,, \nonumber\\
\hat{H}_{+-;-1,1} & = - \hat{H}^*_{-+;1,-1}  = \, 2\,\sqrt{3}\,e^2\,e_q^2\,e^{-i2\phi_q}\,\frac{\hat{s}}{\hat{s}+Q^2}\,\sqrt{\frac{\hat{t}}{\hat{u}}}  =  \,\sqrt{3}\,e^2\,e_q^2\,e^{-i2\phi_q}\,\frac{\xi-x_B}{\xi}\,\sqrt{\,\frac{\zeta}{1-\zeta}\,}\,, \nonumber\\
\hat{H}_{+-;-1,-1} & = - \hat{H}_{-+;1,1}   = \,  2\,\sqrt{3}\,e^2\,e_q^2\,\frac{Q^2}{\hat{s}+Q^2}\,\sqrt{\frac{\hat{t}}{\hat{u}}}  = \, 2\,\sqrt{3}\,e^2\,e_q^2\,\frac{x_B}{\xi}\,\sqrt{\,\frac{\zeta}{1-\zeta}\,}\,, \nonumber\\
\hat{H}_{+-;0,\pm 1} & = - \hat{H}_{-+;0,\pm 1}  = \, \pm 2\,\sqrt{6}\,e^2\,e_q^2\,e^{\mp i\phi_q}\,\frac{\sqrt{\hat{s}}Q}{\hat{s}+Q^2}  = \, \pm 2\,\sqrt{6}\,e^2\,e_q^2\,e^{\mp i\phi_q}\,\frac{\sqrt{\,x_B(\xi-x_B)}}{\xi}\,. 
\label{eq:H-all}
\end{align}

%\bibliography{sample}
%\bibliographystyle{utphys}  

\providecommand{\href}[2]{#2}\begingroup\raggedright\endgroup

\end{document}